\documentclass{ws-procs9x6}

\begin{document}

\title{Phase Conversion After a Chiral Transition: 
Effects from Inhomogeneities and Finite Size\thanks{\uppercase{P}resented 
at \uppercase{S}trong and \uppercase{E}lectroweak \uppercase{M}atter 2004, 
\uppercase{H}elsinki, \uppercase{J}une 16-19, 2004}}

\author{Eduardo S. Fraga}

\address{Instituto de F\'\i sica, Universidade Federal do Rio de 
Janeiro \\
C.P. 68528, Rio de Janeiro, RJ 21941-972, Brazil\\
E-mail: fraga@if.ufrj.br}

\maketitle

\abstracts{
We discuss the kinetics of phase conversion, through the nucleation
of bubbles and spinodal decomposition, after a chiral transition
within an effective field theory approach to low-energy QCD. 
We study possible effects resulting from the
finite size of the expanding system for both the initial and the
late-stage growth of domains, as well as those effects due to
inhomogeneities in the chiral field which act as a background for the
fermionic motion.}



To model the mechanism of chiral symmetry breaking present 
in QCD, and to
study the dynamics of phase conversion after a temperature-driven chiral
transition, one can resort to low-energy effective models. 
In particular, to study the mechanisms of bubble nucleation and
spinodal decomposition in a hot expanding plasma,
it is common to adopt the linear $\sigma$-model
coupled to quarks, where the latter comprise the
hydrodynamic degrees of freedom of the 
system\cite{ove,Scavenius:1999zc,Scavenius:2000qd,Scavenius:2001bb,paech,Aguiar:2003pp}.
The gas of quarks provides a thermal bath in which the long-wavelength
modes of the chiral field evolve, and the latter plays the role of an order
parameter in a Landau-Ginzburg approach to the description of the chiral
phase transition. The gas of quarks
and anti-quarks is usually treated as a heat bath for the chiral field,
with temperature $T$. The standard procedure is then integrating over
the fermionic degrees of freedom, using a classical approximation
for the chiral field, to obtain a formal expression for the thermodynamic
potential of an infinite system.



Let us consider a chiral field $\phi=(\sigma,\vec{\pi})$, 
where $\sigma$ is
a scalar field and $\pi^i$ are pseudoscalar fields playing the role of the
pions, coupled to two flavors of quarks according to the Lagrangian
\begin{equation}
\mathcal{L} = \overline{q}[i\gamma^{\mu}\partial _{\mu} + \mu_q\gamma^0 -
M(\phi)]q + \frac{1}{2}\partial_{\mu}\phi \partial^{\mu}\phi - V(\phi)\; .
\end{equation}
Here $q$ is the constituent-quark field $q=(u,d)$ and $\mu_q=\mu/3$ is the
quark chemical potential. The interaction between the quarks and the chiral
field is given by 
$M(\phi) = g \left(\sigma + i\gamma_5 \vec{\tau}\cdot \vec{\pi}\right)$ ,
and 
$V(\phi)= \frac{\lambda^2}{4}
\left(\sigma^2 + \vec{\pi}^2 -v^2\right)^2 - h_q\sigma$ 
is the self-interaction potential for $\phi$. 

The parameters above are
chosen such that chiral $SU_{L}(2) \otimes SU_{R}(2)$ symmetry is
spontaneously broken in the vacuum. The vacuum expectation values of the
condensates are
$\langle\sigma\rangle =\mathit{f}_{\pi}$ and $\langle\vec{\pi}\rangle =0$,
where $\mathit{f}_{\pi}=93$~MeV is the pion decay constant.
The explicit symmetry breaking term is due to the finite current-quark
masses and is determined by the PCAC relation, giving
$h_q=f_{\pi}m_{\pi}^{2}$, where $m_{\pi}=138$~MeV is the pion mass. This
yields $v^{2}=f^{2}_{\pi}-{m^{2}_{\pi}}/{\lambda ^{2}}$. The value of
$\lambda^2 = 20$ leads to a $\sigma$-mass,
$m^2_\sigma=2\lambda^{2}f^{2}_{\pi}+m^{2}_{\pi}$, equal to 600~MeV.
For $g>0$, the finite-temperature one-loop effective potential also
includes a contribution from the quark fermionic determinant.
In what follows, we treat the gas of quarks as a heat
bath for the chiral field, with temperature $T$ and baryon-chemical
potential $\mu$. Then, one can integrate over the fermionic degrees
of freedom, obtaining an effective theory for the chiral field $\phi$.
Using a classical approximation for the chiral
field, one obtains the thermodynamic potential 
\begin{equation}
\Omega(T,\mu,\phi) = V(\phi) - \frac{T}{\mathcal{V}}
\ln\det\{[G_E^{-1}+W(\phi)] / T \} \; ,
\end{equation}
where $G_E$ is the fermionic Euclidean propagator. 
From the thermodynamic potential one can obtain
all the thermodynamic quantities of interest. 



To compute correlation functions and thermodynamic
quantities, one has to evaluate the fermionic determinant within some 
approximation scheme. In $1D$ systems
one can usually resort to exact analytical methods\cite{Fraga:1994xd}.
In practice, however, the determinant is usually calculated to one-loop
order assuming a homogeneous and static background chiral
field. Nevertheless, for a system that is in the process
of phase conversion after a chiral transition, one expects inhomogeneities
in the chiral field to play a role in driving the system to the 
true ground state.

We propose an approximation procedure to evaluate
the finite-temperature fermionic determinant in the presence
of a chiral background field, which systematically incorporates
effects from inhomogeneities in the chiral field through a
derivative expansion. The method is valid for the case in which
the chiral field varies smoothly, and allows one to extract information
from its long-wavelength behavior, incorporating corrections order
by order in the derivatives of the field.

The Euler-Lagrange equation for static chiral field configurations
contains a term which represents the fermionic density 
$\rho(\vec{x}_0)=(\nu_q/\mathcal{V}) \left\langle\vec{x}_0
\left\vert (G_E^{-1}+M(\hat{x}))^{-1}\right\vert
\vec{x}_0\right\rangle$ ,
where $\vert\vec{x}_0\rangle$ is a position eigenstate with eigenvalue
$\vec{x}_0$, and $\nu_q=12$ is the color-spin-isospin degeneracy factor. 
In momentum representation:
\begin{equation}
\rho(\vec{x}_0)=\nu_q
T\sum_n \int \frac{d^3k}{(2\pi)^3}
e^{-i\vec{k}\cdot\vec{x}} 
\frac{1}{\gamma^0 (i\omega_n+\mu)-\vec{\gamma}\cdot\vec{k}+M(\hat{x})}
e^{i\vec{k}\cdot\vec{x}} \; .
\end{equation}
We can transfer the $\vec{x}_0$ dependence to $M(\hat{x})$ through a
unitary transformation, expand $M(\hat{x}+\vec{x}_0)$ around $\vec{x}_0$, 
and use $\hat{x}^i=-i\nabla_{k_i}$ to write
\begin{eqnarray}
&&\rho(\vec{x}_0)=\nu_q T\sum_n \int \frac{d^3k}{(2\pi)^3}
\frac{1}{\gamma^0(i\omega_n+\mu)-\vec{\gamma}\cdot\vec{k}+M(\vec{x}_0)}
\times\nonumber \\
&&\left[ 1+ \Delta M(-i\nabla_{k_i},\vec{x}_0)
\frac{1}{\gamma^0(i\omega_n+\mu)-\vec{\gamma}\cdot\vec{k}+M(\vec{x}_0)}
\right]^{-1}\, ,
\label{brackets}
\end{eqnarray}
where $\vec{x}_0$ is a c-number, not an operator, and 
$\Delta M(\hat{x},\vec{x}_0)=\nabla_i M(\vec{x}_0)\hat{x}^i +
\frac{1}{2}\nabla_i \nabla_j M(\vec{x}_0)\hat{x}^i \hat{x}^j +
\cdots$.

If we focus on the long-wavelength properties
of the chiral field and assume that the
static background, $M(\vec{x})$, varies smoothly and fermions
transfer a small ammount of momentum to the chiral field,  
we can expand the expression above in a power series:
\begin{eqnarray}
&&\rho(\vec{x})=\nu_q T\sum_n \int \frac{d^3k}{(2\pi)^3}
\frac{1}{\gamma^0(i\omega_n+\mu)-\vec{\gamma}\cdot\vec{k}+M(\vec{x})}
\times \nonumber \\
&&\sum_\ell (-1)^\ell
\left[ \Delta M(-i\nabla_{k_i},\vec{x})
\frac{1}{\gamma^0(i\omega_n+\mu)-\vec{\gamma}\cdot\vec{k}+M(\vec{x})}
\right]^{\ell}\, ,
\label{expansion}
\end{eqnarray}
\begin{equation}
\Delta M(-i\nabla_{k_i},\vec{x})=
\nabla_i M\left(\frac{1}{i}\right)\nabla_{k_i}+
\frac{1}{2}\nabla_i \nabla_j M\left(\frac{1}{i}\right)^2
\nabla_{k_i}\nabla_{k_j} + \cdots \, ,
\end{equation}
which provides a systematic procedure to incorporate 
corrections brought about
by inhomogeneities in the chiral field to the quark density, so that
we can calculate
$\rho(\vec{x})=\rho_0(\vec{x})+\rho_1(\vec{x})+\rho_2(\vec{x})+\cdots$
order by order in powers of the derivative of the background, $M(\vec{x})$.
The leading-order term in this gradient expansion for $\rho(\vec{x})$
can be easily calculated and yields the well-known mean field result 
for $\rho$. The net effect
of this leading term is to correct the potential for the chiral
field to $V_{eff}= V(\phi)+V_q(\phi)$, where
\begin{equation}
V_q\equiv -\nu_q T \int \frac{d^3k}{(2\pi)^3}
\ln\left( e^{[E_k(\phi)-\mu_q]/T}+1 \right)+
+ (\mu_q \to -\mu_q) \; ,
\end{equation}
where $E_k(\phi)=\sqrt{\vec{k}^2+M(\phi)}$.
This sort of effective potential is commonly used as the
thermodynamic potential in a phenomenological
description of the chiral transition for an expanding
quark-gluon plasma created in a high-energy heavy-ion 
collision\cite{Scavenius:1999zc,Scavenius:2000qd,Scavenius:2001bb,paech,Aguiar:2003pp}. 
However, the presence of a non-trivial
background field configuration, {\it e.g.} a bubble, can in
principle dramatically modify the Dirac spectrum,
hence the determinant\cite{Fraga:1994xd,polarons}. 
Results for the correction of the 
Laplacian term will be presented elsewhere\cite{prep}.



In the process of phase conversion through bubble nucleation 
in a QGP of finite size, the set of all supercritical bubbles 
integrated over time will eventually drive the entire system 
to its true vacuum. The scales that determine the importance 
of finite-size effects are the typical linear size of the system,
the radius of the critical bubble and the correlation length. 
For definiteness, let us assume our system is described by
a coarse-grained Landau-Ginzburg potential, $U(\phi,T)$, whose 
coefficients depend on the temperature. For the case to be considered,
the scalar order parameter, $\phi$, is {\it not} a conserved quantity, and
its evolution is given by the time-dependent Landau-Ginzburg equation
\begin{equation}
\frac{\partial\phi}{\partial t}=
\gamma \left[ \nabla^2\phi - U'(\phi,T) \right]\quad ,
\label{reaction}
\end{equation}
where $\gamma$ is the response coefficient which defines a time scale for
the system. The equation above is a standard reaction-diffusion
equation, and describes the approach to equilibrium.

If $U(\phi,T)$ is such that it allows for the 
existence of bubble solutions
(taken to be spherical for simplicity), then supercritical (subcritical)
bubbles expand (shrink), in the thin-wall limit, with the following
velocity:
\begin{equation}
\frac{dR}{dt}=\gamma (d-1) \left[ \frac{1}{R_c}-\frac{1}{R(t)}  \right]
\quad ,
\label{allen-cahn}
\end{equation}
where $R_c=(d-1)\sigma/\Delta F$ and
$\Delta F$ is the difference in free energy
between the two phases. 
The equation above relates the velocity of a domain wall to the local
curvature. The response coefficient, $\gamma$, can be related to some
characteristic collision time.
One can measure the importance of finite-size effects for
the case of heavy-ion collisions by comparing, for instance, the
asymptotic growth velocity ($R>>R_c$)
for nucleated hadronic bubbles to the
expansion velocity of the plasma. In the Bjorken picture,  
the typical length scale of the expanding system is 
$L(T)\approx (v_z t_c)(T_c/T)^3=
L_0 (T_c/T)^3$, 
where $v_z$ is the collective fluid velocity 
and $L_0\equiv L(T_c)$ is the initial linear scale of the
system for the nucleation process which starts at $T\leq T_c$.

The relation between time and temperature provided by the
cooling law that emerges from the Bjorken picture suggests
the comparison between the following ``velocities'':
\begin{equation}
v_b\equiv \frac{dR}{dT}=
-\left(\frac{3b\ell L_0}{2v_z\sigma T_c^2}\right)
\left(\frac{T_c}{T}\right)^5 \left( 1-\frac{T}{T_c} \right)\quad ,
\end{equation}
the asymptotic bubble growth ``velocity'', and the plasma
expansion ``velocity'' 
$v_L\equiv (dL/dT)=-(3L_0/T_c)(T_c/T)^4$.
The quantity $b$ is a number of order one to first approximation,
and comes about in the estimate of the phenomenological
response coefficient $\gamma (T)\approx b/2T$. 
Using the numerical values adopted previously and
$\sigma/T_c^3\sim 0.1$, we obtain\cite{finite2}
\begin{equation}
\frac{v_b}{v_L}\approx \frac{20}{v_z}
\left( \frac{T_c}{T} -1\right)\quad .
\end{equation}

One thus observes that the bubble growth velocity becomes
larger than the expansion velocity for a supercooling of
order $\theta\approx v_z/20 \leq 5\%$. A simple estimate
points to a critical radius larger than $1~$fm at
such values of supercooling\cite{Scavenius:2001bb}.
Therefore, finite-size effects appear to be an important
ingredient in the phase conversion process right from the
start in the case of high-energy heavy-ion 
collisions\cite{finite2}.




\section*{Acknowledgments}

Part of this paper is based on work done in collaboration 
with R. Venugopalan. 
E.S.F. is partially supported by CAPES, CNPq, FAPERJ and FUJB/UFRJ.


\end{document}